\documentclass[manuscript,acmsmall,nonacm]{acmart}

\usepackage{graphicx}
\usepackage{multirow}
\usepackage{graphicx}
\usepackage{graphics}      
\usepackage{color}
\usepackage{booktabs}
\usepackage{float}
\usepackage[export]{adjustbox}
\usepackage{soul}

\usepackage{caption,subcaption,graphicx}
\definecolor{OliveGreen}{rgb}{0,0.6,0}
\definecolor{lightred}{rgb}{1, 0.8, 0.8}
\definecolor{lightgreen}{rgb}{0.8, 1, 0.8}
\definecolor{blue}{rgb}{0, 0, 0}

\usepackage{color-edits}
\addauthor{ml}{olive} 
\addauthor{cj}{purple} 

\AtBeginDocument{%
  }

\begin{document}

\title[Understanding the Effect of Counterfactual Explanations on Trust and Reliance on AI]{Understanding the Effect of Counterfactual Explanations on Trust and Reliance on AI for Human-AI Collaborative Clinical Decision Making}

\author{Min Hun Lee}
\email{mhlee@smu.edu.sg}
\author{Chong Jun Chew}
\email{cjchew.2020@scis.smu.edu.sg}
\affiliation{%
  \institution{Singapore Management University}
  \city{Singapore}
  \country{Singapore}
}

\renewcommand{\shortauthors}{Lee and Chew}

\begin{abstract}
Artificial intelligence (AI) is increasingly being considered to assist human decision-making in high-stake domains (e.g. health). However, researchers have discussed an issue that humans can over-rely on wrong suggestions of the AI model instead of achieving human AI complementary performance. In this work, we utilized salient feature explanations along with what-if, counterfactual explanations to make humans review AI suggestions more analytically to reduce overreliance on AI and explored the effect of these explanations on trust and reliance on AI during clinical decision-making. We conducted an experiment with seven therapists and ten laypersons on the task of assessing post-stroke survivors' quality of motion, and analyzed their performance, agreement level on the task, and reliance on AI without and with two types of AI explanations. Our results showed that the AI model with both salient features and counterfactual explanations assisted therapists and laypersons to improve their performance and agreement level on the task when `right' AI outputs are presented. While both therapists and laypersons over-relied on `wrong' AI outputs, counterfactual explanations assisted both therapists and laypersons to reduce their over-reliance on `wrong' AI outputs by 21\% compared to salient feature explanations. Specifically, laypersons had higher performance degrades by 18.0 f1-score with salient feature explanations and 14.0 f1-score with counterfactual explanations than therapists with performance degrades of 8.6 and 2.8 f1-scores respectively. Our work discusses the potential of counterfactual explanations to better estimate the accuracy of an AI model and reduce over-reliance on `wrong' AI outputs and implications for improving human-AI collaborative decision-making.
\end{abstract}




\maketitle

\section{Introduction}
As advanced artificial intelligence (AI) and machine learning (ML) models have achieved equivalent results or outperformed humans at decision-making tasks (e.g. screening lung cancer \cite{ardila2019end}), these AI and ML models are increasingly being considered to increase efficiency and reduce the cost of performing decision-making tasks from various types of organizations and domains \cite{stone2016artificial} (e.g. health \cite{rajpurkar2022ai,lee2021human,cai2019human}, bail decisions \cite{kleinberg2018human}, child welfare services \cite{brown2019toward}, university admissions decisions \cite{cheng2019explaining}, etc.). Specifically, researchers have discussed the potential of human and AI/ML teaming to achieve better results than either humans or AI/ML models alone \cite{bansal2019beyond,lee2021human,lee2022towards}. However, previous research works have discussed that users might place too much trust in the AI/ML system and even agree with `wrong' AI outputs \cite{bussone2015role,buccinca2021trust,lai2019human}.

Many researchers have discussed that the explainability \cite{tjoa2020survey} of a system is critical for human-AI collaborative decision-making \cite{cai2019human,lee2021human}. In particular, humans can review AI explanations to understand how the AI models generate outputs \cite{tjoa2020survey} and identify whether an AI output is right or not \cite{cai2019human,lee2021human}. There is growing number of studies to evaluate the effect of AI explanations on decision-making tasks \cite{lai2019human,lee2021human,cai2019human,bansal2019beyond,bussone2015role,wang2021explanations,buccinca2021trust}. For instance, researchers studied the effect of AI explanations on over-reliance using the simulated AI models or the tasks that do not require domain experts, such as judges or clinicians \cite{lai2019human,wang2021explanations, bansal2019beyond}. However, there has been contradictory perspectives on the effect of AI explanations on user's trust on an AI output: users' trust in an algorithmic decision is not affected by the explanation interface \cite{cheng2019explaining} or can be increased by just presenting explanations \cite{bansal2019beyond}. 

In this work, we contribute to an empirical study that analyzes the effect of AI explanations on users' trust and reliance on AI during clinical decision-making. Specifically, we focus on the task of assessing post-stroke survivors' quality of motion \cite{lee2020co}. Among various types of AI explanations \cite{tjoa2020survey,lee2023exploring}, this work explores salient feature analysis and counterfactual explanations for the following reasons. First, the previous research describes that therapists preferred to review feature-based explanations on rehabilitation assessment tasks \cite{lee2020co}. However, the previous research discusses the issues of these feature-based explanations on overtrust in AI \cite{kaur2020interpreting,wang2021explanations,lai2021towards}. One potential reason for overreliance on AI might be that humans mostly employ heuristics and shortcuts while rarely involving analytical thinking during decision-making \cite{kahneman2011thinking,buccinca2021trust}. Previous research \cite{buccinca2021trust} discusses the potential of cognitive forcing functions (e.g. not showing AI suggestions by default or waiting before showing AI suggestions) to increase analytical thinking and reduce overreliance. In this work, we assume that reviewing counterfactual explanations \cite{mothilal2020explaining,verma2020counterfactual,wang2021explanations} will allow a user to critically think of how to change an AI output and more rigorously review an AI output than widely used AI explanations (e.g. feature-based or example-based explanations) that show information relevant to an AI output. We hypothesize that reviewing counterfactual explanations \cite{mothilal2020explaining,verma2020counterfactual,wang2021explanations} will improve the user's analytical review of an AI output, assist the user to achieve better calibrated trust in AI, and reduce overreliance on it.

To this end, we conduct a within-subject experiment with seven therapists and ten laypersons to compare the effect of counterfactual explanations with one of the widely used AI explanations, salient feature explanations \cite{lai2019human,cheng2019explaining,lee2021human,bansal2019beyond}. Our results show that the human + AI team with both salient feature and counterfactual explanations improved the performance and agreement level on decision-making tasks only when `right' AI outputs were presented. In contrast, when `wrong' AI outputs were presented, the human + AI team with salient feature analysis had higher overreliance on `wrong' AI outputs while the human + AI team with counterfactual explanations reduced overreliance on `wrong' AI outputs by 21\% compared to salient feature explanations. 

When we analyzed the performance and the effect of AI explanations by therapists and laypersons, therapists had lower performance degradation and overreliance on `wrong' AI outputs than laypersons: therapists' human + AI team performance was lower than their human alone performance by 8.6 f1-score with salient feature and 2.8 f1-score with counterfactual explanations; laypersons' human + AI team performance was lower than their human alone performance by 18.0 f1-score with salient feature and 14.0 f1-score with counterfactual explanations. Overall, reviewing counterfactual explanations assisted both therapists and laypersons to diversify their assessment (i.e. lower agreement level) and have more cases of rejecting `wrong' AI outputs and fewer cases of agreeing with `wrong' AI outputs than salient feature analysis by 19\% from therapists and by 35\% from laypersons. 

When it comes to a self-reported usability score, therapists and laypersons had a higher self-reported trust score (73.78 out of 100) on the AI system with salient feature analysis than the AI system with counterfactual explanations (45.20 out of 100). The self-reported trust score of the system with counterfactual explanations is closer to the system performance (0.375: 3 right outputs of out 8) than that of the system with salient feature analysis.

Overall, this work provides new insights into the potential of counterfactual explanations to reduce overreliance on `wrong' AI outputs and better estimate the performance of an AI model through a user study using uncontrolled AI outputs and explanations with therapists and laypersons on clinical decision-making tasks (i.e. rehabilitation assessment). In addition, our work compares the effect of AI outputs and explanations on domain experts and lay group participants. Our work advances ongoing discussions around the implications for improving human-AI collaborative decision-making in high-stake domains (e.g. health) \cite{letham2015interpretable,cai2019hello,lee2021human}.

\section{Related Work}
\subsection{Towards Human-AI Collaborative Decision-Making}
With the recent advance in artificial intelligence (AI) and machine learning algorithms, AI/ML models are increasingly being considered to assist humans' decision-making tasks in a variety of domains (e.g. health). Instead of applying fully autonomous AI systems, researchers have explored the feasibility of human-AI collaborative decision-making, in which an AI model provides humans new data-driven insights on a task for achieving complementary performance, outperforming neither of the AI or the human alone \cite{bansal2019beyond,lee2021human,lee2022towards,kawakami2022improving}.
For instance, a deep learning-based system has been used in clinics to bring new data-driven insights to assist the diagnosis of cancer \cite{cai2019hello}, the detection of diabetic eye disease \cite{beede2020human}, or the assessment of physical stroke rehabilitation assessment \cite{lee2021human}.

Although previous research describes the potential of AI/ML systems to outperform human experts on prediction tasks \cite{esteva2017dermatologist,lee2019learning,kleinberg2018human}, it still remains a challenge to develop and integrate these systems in practice due to the lack of human-centered designs and performing as a "black-box" system  \cite{khairat2018reasons,yang2019unremarkable,cabitza2017unintended,cai2019human,cai2019hello,lee2020co,green2019principles}.
For the issue of lack of human-centered designs, there has been increasing recent research efforts \cite{sendak2020human,yang2019unremarkable,green2019principles,beede2020human,lee2020co,cai2019hello} that highlight the importance of involving stakeholders to understand their practices and needs \cite{yang2019unremarkable,cai2019hello,lee2020co} and socio-environmental factors \cite{beede2020human} for the design and evaluation of a system. For instance, Yang et al. \cite{yang2019unremarkable} conducted a field evaluation on the design of a decision support tool for cardiologists with synthetic data and found that clinicians are more likely to embrace a tool that augments their decision-making in natural and intuitive ways. Lee et al. \cite{lee2020co} conducted interviews and focus-group sessions with therapists to understand the challenges and needs during rehabilitation assessment to design a human-centered decision support system.

\subsection{User Studies of Explainable AI}
In addition, researchers have discussed the importance of an AI explanation to communicate an AI output to a user \cite{bansal2019beyond,feng2019can,kaur2020interpreting} and realize human-AI collaboration \cite{mueller2019explanation} for a decision-making task \cite{cai2019hello,lee2021human}. 
There has been a growing number of studies that have evaluated the effect of AI explanations in diverse decision-making tasks (e.g. house price prediction \cite{poursabzi2021manipulating}, image classification \cite{alqaraawi2020evaluating}, student admission \cite{cheng2019explaining}, deception detection \cite{lai2019human}, stroke rehabilitation assessment \cite{lee2021human}) and aspects, such as whether an AI explanation assists a user to debug \cite{kaur2020interpreting} or update an AI model \cite{cai2019human,lee2021human} or improves user's trust in AI \cite{kunkel2019let,cai2019human,buccinca2021trust,panigutti2022understanding}. 

For instance, Alqaraawi et al. \cite{alqaraawi2020evaluating} conducted a user study on image classification tasks to evaluate the performance of the saliency map, an XAI technique that highlights input pixels in the original images that contribute to a model prediction and discussed the limited usefulness of saliency map to assist participants to anticipate a model output. Wang and Yin \cite{wang2021explanations} conducted the randomized experiment using four types of common model-agnostic explainable AI methods and discussed the effect of AI explanations could be largely different where people have varying levels of domain expertise. 

There have been contradictory perspectives on the effect of AI explanations on user's trust in an AI output: users' trust in an algorithmic decision is not affected by the explanation interface \cite{cheng2019explaining} or can be increased by just presenting explanations \cite{bansal2019beyond} or even when explanations are randomly chosen \cite{lai2019human}. According to the study with MTruk worker on the task of deception detection task \cite{lai2019human}, Lai and Tan discussed that the presentation of AI-predicted labels and explanations improves human performance on a task. In contrast, Bussone et al. discussed that providing richer explanations could lead to a harmful effect: overreliance on the system \cite{bussone2015role}. Along this issue, Buccinca et al. \cite{buccinca2021trust} discussed the cognitive forcing intervention, such as slowing down the process and asking the person to make a decision before seeing the AI recommendation, can reduce the overreliance on AI. 

In addition to the contradictory perspectives on the effect of explanations on trust, our research community still requires additional studies to understand the effect of AI explanations on overreliance \cite{parasuraman1997humans}. Specifically, even if Bussone et al. \cite{bussone2015role} and Buccinca et al. \cite{buccinca2021trust} investigated the effect of AI explanations on user's overreliance, previous research utilized a mock-up decision support system that operates with the wizard-of-oz approach \cite{bussone2015role} or a simulated AI model \cite{buccinca2021trust}. Other works that utilize AI/ML models focused on tasks that do not require domain experts, such as judges or clinicians \cite{lai2019human,wang2021explanations, bansal2019beyond}. 

In this work, we focused on the AI-assisted clinical decision-making task (i.e. physical stroke rehabilitation assessment) and investigate the effect of the salient feature and counterfactual explanations on users' trust and overreliance on AI. Specifically, this work utilized uncontrolled AI model outputs and explanations implemented by the dataset of 15 post-stroke survivors in contrast to existing previous research that utilizes simulated and controlled AI outputs and explanations \cite{bussone2015role,buccinca2021trust,naiseh2023different} to understand the effect of AI explanations and the issue of overreliance. This work contributes to increasing knowledge on the effect of AI explanations by (i) comparing human alone and human + AI team in terms of performance, agreement level, and the number of `right' or `wrong' decisions and (ii) analyzing these evaluation metrics between domain experts (i.e. therapists) and laypersons. This work further discusses the potential of counterfactual explanations as a cognitive forcing function to better achieve a calibrated trust in AI and reduce overreliance on AI and implications for improving human-AI collaborative decision-making in high-stake domains (e.g. health) \cite{letham2015interpretable,cai2019hello,lee2021human}.

\section{Study Design}
\begin{figure*}[!thp]
\centering
\begin{subfigure}[t]{.44\columnwidth}
\centering
  \includegraphics[width=1.0\linewidth]{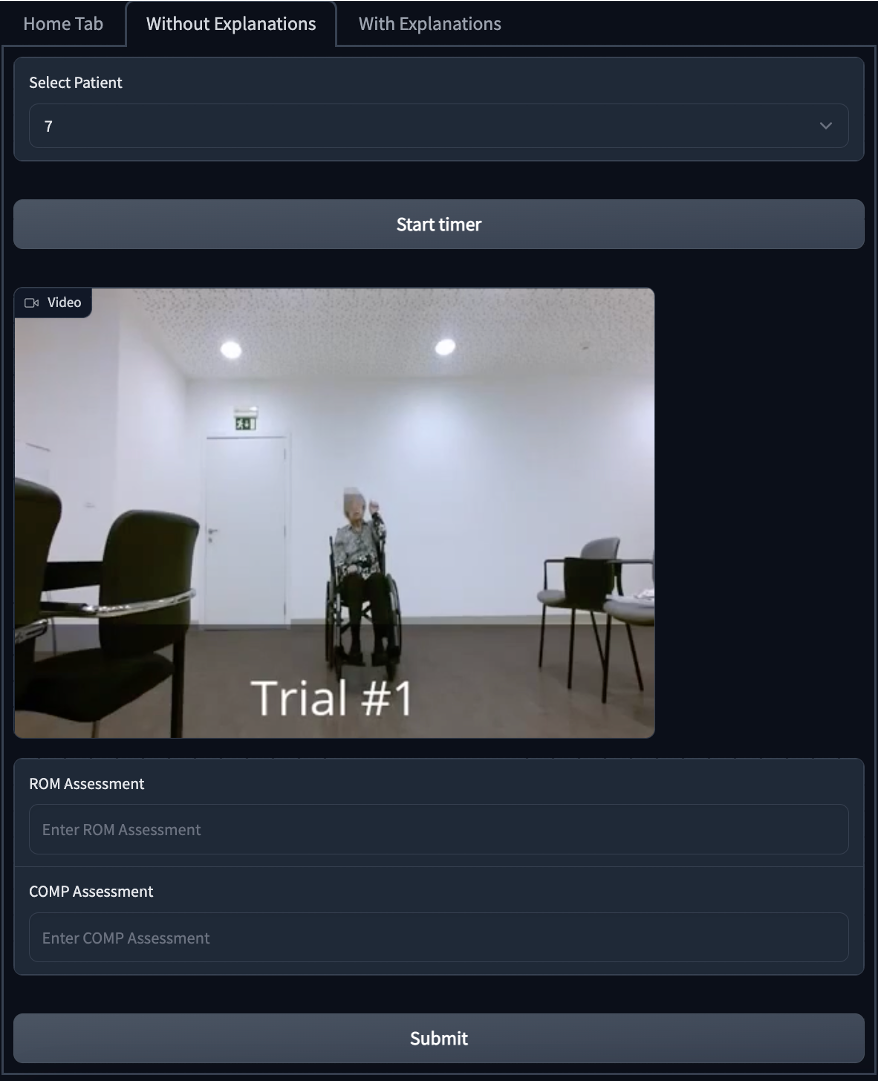}
  \caption{}
  \label{fig:IntVideo}
\end{subfigure}
\begin{subfigure}[t]{.455\columnwidth}
  \centering
  \includegraphics[width=1.0\linewidth]{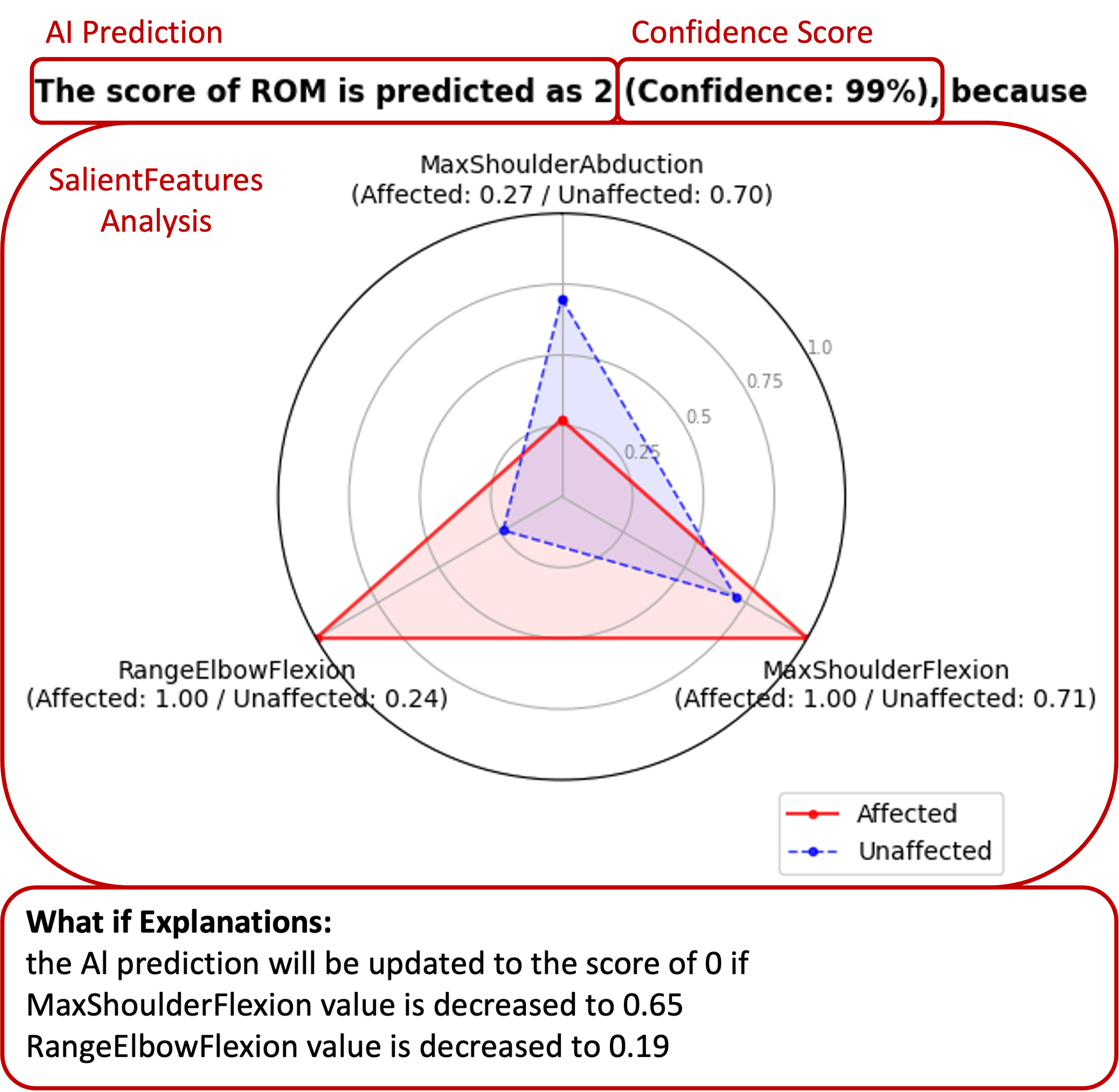}
  \caption{}
  \label{fig:IntfAIOutputs}
\end{subfigure}
\caption{The AI-based decision support system that presents (a) the video of post-stroke survivor's exercises and (b) the predicted assessment score by AI along with salient feature-based explanations that compare a post-stroke survivor's unaffected and affected side using the top three most important features and counterfactual, what-if explanations that describe how input features need to be changed to flip an AI output (e.g. `correct' to `incorrect' ROM).}\label{fig:interface}
\end{figure*}

The primary research question of this work is to investigate the effect of AI explanations on users' trust and reliance on imperfect AI outputs. Building upon growing works on the usage of explainable AI methods for improving AI-assisted decision-making \cite{bussone2015role,buccinca2021trust,wang2021explanations}, we hypothesize that counterfactual explanations \cite{mothilal2020explaining,verma2020counterfactual}, a type of AI explanations that describe how the inputs can be modified to achieve an AI output in a certain way, will increase user's analytic reviews and deliberations on an AI output and reduce user's overreliance on `wrong' AI outputs. To this end, we conducted a within-subject experiment with therapists and laypersons in the context of assessing post-stroke survivors' quality of motion. Specifically, we compared the effect of using a decision support system with counterfactual explanations (Figure \ref{fig:interface}) to a baseline system with one of the widely used explainable AI techniques, salient features, calculating the importance of input features,  \cite{lipton2018mythos,NIPS2017Shap,lee2020co}. Our study aims to explore the following research question:

\begin{itemize}
    \item How do counterfactual explanations impact user's (1) performance \& agreement level on decision-making tasks and (2) reliance and trust on AI outputs?
\end{itemize}

\subsection{Clinical Decision Making Task: Physical Stroke Rehabilitation Assessment}
\begin{figure}[tp!]
\centering
\begin{subfigure}[t]{.25\columnwidth}
\centering
  \includegraphics[width=.68\columnwidth]{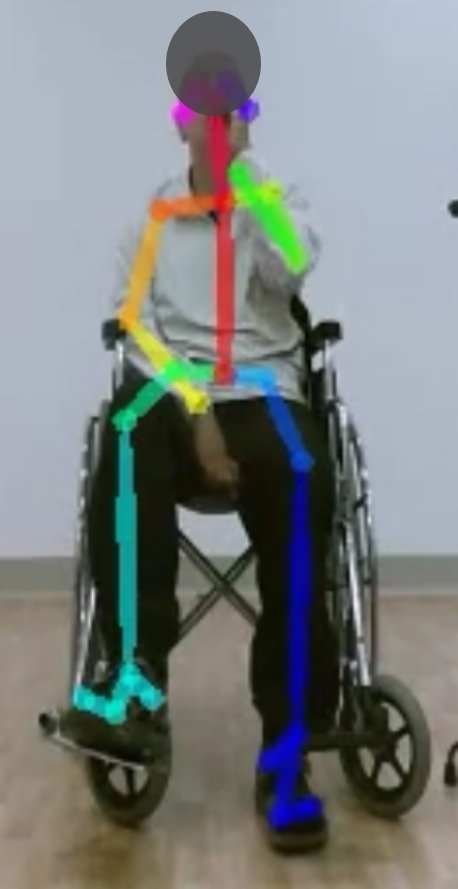}
  \caption{E1-Unaffected}
  \label{fig:sample-compensation-normal-e1-p14}
\end{subfigure}
\begin{subfigure}[t]{.25\columnwidth}
\centering
  \includegraphics[width=.69\columnwidth]{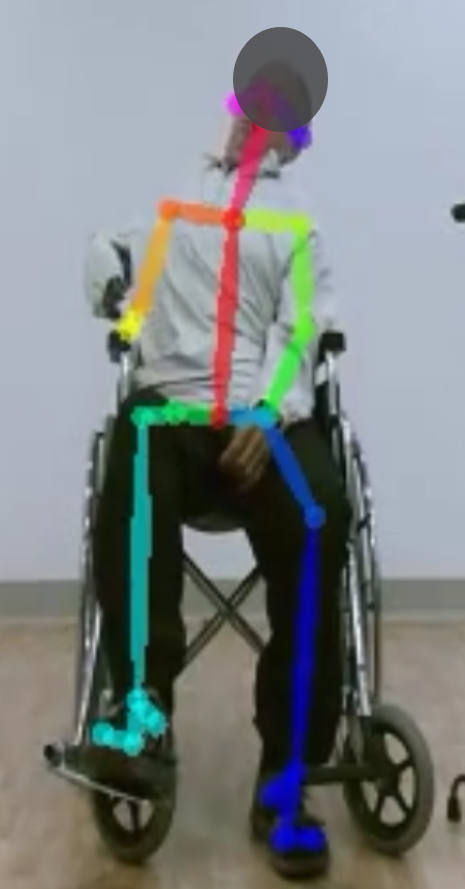}
  \caption{E1-Affected}
  \label{fig:sample-compensation-abnormal-e1-p14}
\end{subfigure}
\caption{Sample Unaffected and Affected Motions of an Exercise: (a) a patient can raise the patient's wrist to the mouth, (b) a patient compensated with trunk and shoulder joints.}~\label{fig:sample-exercise}
\end{figure}

In this work, we focus on a clinical decision-making task: assessing the quality of motion of patients affected by stroke, the second leading cause of death and third most common contributor to disability \cite{feigin2017global}. Building upon previous works on AI-assisted decision-making on physical stroke rehabilitation assessment \cite{lee2021human}, this work utilizes an upper-limb rehabilitation exercise (Figure \ref{fig:sample-exercise}) and two performance components of rehabilitation assessment: Range of Motion (ROM) and Compensation. 

For an exercise, a post-stroke survivor has to raise his or her wrist to the mouth as if drinking water (Figure \ref{fig:sample-compensation-normal-e1-p14}). For the rehabilitation assessment, the \textit{`ROM'} component refers to how closely a post-stroke survivor achieves the target position of an exercise (e.g. bring the wrist to the mouth) and the \textit{`Compensation'} component indicates whether a post-stroke survivor involves any unnecessary joints to perform an exercise (e.g. leaning trunk to the side and backward - Figure \ref{fig:sample-compensation-abnormal-e1-p14}).

For the rehabilitation assessment task, participants went through the tutorial on rehabilitation assessment and were asked to review the video of post-stroke survivor's exercises and assess the post-stroke survivor's quality of motion in terms of the \textit{`ROM'} and  the \textit{`Compensation'}. The score guidelines for rehabilitation assessment can be found in Table \ref{tab:score-guidelines} in the Appendix.

\subsection{Dataset, AI Models and Explanations}
\subsubsection{Dataset and Kinematic Features}
\hfill\\
We utilized the dataset of a \textit{`Bring a cup to the mouth'} upper-limb exercise from 15 post-stroke survivors \cite{lee2019learning}. Specifically, this dataset includes (1) the 300 videos of 15 post-stroke survivors performing the exercise (10 trials using their unaffected and affected side by stroke respectively), (2) their estimated joint positions using a Kinect sensor v2, and (3) the annotations by the expert therapist, who evaluated the post-stroke survivors' using clinically validated Fugl Meyer Assessment \cite{sullivan2011fugl} and watched the recorded videos without reviewing any AI outputs.

Given the estimated joint positions of post-stroke survivors' exercises, we extracted various kinematic features based on the previous work \cite{sullivan2011fugl,lee2019learning}. For the \textit{`ROM'} component, we extracted joint angles (e.g. elbow flexion, shoulder flexion, elbow extension), normalized relative trajectory (i.e. Euclidean distance between two joints - head and wrist, head and elbow), and normalized trajectory distance (i.e. the absolute distance between two joints - head and wrist, shoulder and wrist) in the x, y, z coordinates \cite{lee2019learning}. For the \textit{`Compensation'} component, we extracted normalized trajectories (distances between joint positions of head, spine, and shoulder in the x, y, z coordinates from the initial to current frames) to distinguish the occurrence of a compensated movement \cite{sullivan2011fugl,lee2019learning}

\subsubsection{AI Models and Explanations}\label{sect:methods_mlmodels}
\hfill\\
We utilized a feed-forward Neural Network model to classify the quality of post-stroke survivor's motion due to its out-performance shown in the previous work \cite{lee2019learning}. Specifically, we grid-searched various architectures (i.e. one to three layers with $32, 64, 128, 256, 512$ hidden units) and an adaptive learning rate with different initial learning rates (i.e. $0.0001, 0.005, 0.001, 0.01, 0.1$) using cross-entropy loss and \textit{`AdamOptimizer'} until the tolerance of optimization became $0.0001$ or the maximum $200$ iterations. We applied the leave-one-subject-out cross-validation, which trains a machine learning (ML) model with data from all post-stroke survivors except one post-stroke survivor, and test the model with the held-out post-stroke survivor. The model parameters that achieved the best F1 score during the cross-validation are described in the Appendix. Table \ref{tab:params-ml}. The trained ML model achieved an average F1-score of 0.9285 for the \textit{`ROM'} and an average F1-score of 0.7867 for \textit{`Compensation'} performance component.

After training ML models, we utilized widely used, open-source libraries to generate AI explanations: salient feature analysis and counterfactual explanations. Among various types of AI explanations, this work focuses on exploring salient feature analysis, building upon the previous research that described therapists' preferences in reviewing feature-based explanations on rehabilitation tasks \cite{lee2020co}. However, the previous research describes the issues of these explanations on overtrust in AI \cite{kaur2020interpreting,wang2021explanations,lai2021towards}. This work assumes that counterfactual explanations will induce users to engage in a more critical review of an AI output by thinking about how to change an AI output compared to other widely used AI explanations (e.g. feature-based or example-based explanations) that provide relevant information on confirming an AI output. Thus, this work explores whether the counterfactual, what-if explanations \cite{byrne2019counterfactuals} can assist users to better critically review AI outputs and explanations.

We utilized the SHAP \cite{NIPS2017Shap,roshan2021utilizing} for identifying salient features and the DiCE library \cite{mothilal2020explaining} for generating the counterfactual explanations. 
For salient feature explanations (Figure \ref{fig:IntfAIOutputs}), we identified patient-specific, salient features and utilized only the top three salient features with the highest scores to avoid overwhelming users \cite{lee2020co,kulesza2015principles}. For the presentation of these salient features, we utilized a radar chart to effectively show the comparison of identified features on post-stroke survivors' unaffected and affected sides to follow the therapist's practices \cite{amershi2019guidelines,lee2020co}. For instance, Figure \ref{fig:IntfAIOutputs} shows that the system identified \textit{`MaxShoulderAbduction'}, \textit{`RangeElbowFlexion'}, \textit{`MaxShoulderFlexion'}, statistics of joint angles as the top three, most important features to assess the post-stroke survivor's quality of motion at Figure \ref{fig:IntVideo}. The radar chart describes the differences in identified feature values on post-stroke survivors' unaffected and affected sides. 

The counterfactual explanations describe what changes in feature values lead to updating an AI output in a certain way \cite{byrne2019counterfactuals,guidotti2019factual,mothilal2020explaining}. To generate counterfactual explanations, we applied the model agnostic approach that utilizes the genetic algorithm \cite{olvera2010review,guidotti2019factual,mothilal2020explaining} to find only three counterfactuals close to the query point. In addition, we specified the features to be changed in the DiCE library using the identified salient features by the SHAP library and their desired range using patients' held-out normal data to avoid generating varying and unfeasible explanations.

For the presentation of counterfactual explanations, as we already had a radar chart visualization to describe the comparison between unaffected and affected sides of a post-stroke survivor, we generated textual descriptions of the changes in feature values and AI outputs (Figure \ref{fig:IntfAIOutputs}).
For instance, Figure \ref{fig:IntfAIOutputs} shows that the value of \textit{`MaxShoulderAbduction'} and \textit{`RangeElbowFlexion'} should be reduced to 0.65 and 0.19 respectively to update the output, predicted score of AI from 2 (i.e. full range of motion - ROM) to 0 (i.e. limited ROM).

\subsection{Conditions \& Task Specifications}
\subsubsection{Conditions}
\hfill\\
In this work, we specified two conditions with two different AI explanations to understand their effects on the participants' decision-making task, the rehabilitation assessment of the post-stroke survivors.

\begin{itemize}
    \item The first, baseline condition refers to an AI-based decision support system that presents videos of post-stroke survivor's exercises along with AI prediction scores and salient feature analysis (Figure \ref{fig:IntfAIOutputs} without what-if explanations). 
    \item The second condition refers to the AI system that includes additional counterfactual explanations (Figure \ref{fig:IntfAIOutputs}) compared to the first condition. 
\end{itemize}

In this work, we leverage feature-based explanations for therapists to find evidence and confirm their hypothetical assessment \cite{wang2019designing}. In addition, we explore counterfactual explanations for avoiding therapists' early confirmation \cite{wang2019designing} and inducing more analytical reviews on an AI output to reduce overreliance on AI. As previous work describes therapists preferred to review feature-based explanations to find evidence and confirm their assessment \cite{lee2020co}, we consider that features-based explanations are required by default for therapists to find evidence and confirm their assessment. In addition, counterfactual explanations are required as additional information that serves as a cognitive forcing function to reduce overreliance on AI. Thus, we included both salient feature analysis and counterfactual explanations in the second condition and compared the first and second conditions to understand the effect of counterfactual explanations on users' overreliance on AI.

In the study, we referred to interfaces as ``Condition A'' and ``Condition B'' to avoid biasing participants. We referred to these conditions respectively as the AI with salient features and the AI with counterfactual explanations for clarity throughout the paper. We implemented the web interface of each condition using the Gradio library \cite{abid2019gradio} to conduct the user study. By default, our web interface involves three strategies of cognitive forcing functions \cite{buccinca2021trust} on both Condition A and Condition B to reduce overreliance on AI. Specifically, we implemented the tab menus of `Without Explanations' and 'With Explanations' (Figure \ref{fig:interface}), so that an AI output is not shown to the users from the beginning and allows a user to review AI outputs and explanations and update or confirm their assessment afterward \cite{buccinca2021trust}. In addition, our interface takes around a second to load an AI output and AI explanations instead of explicitly setting 30 seconds of waiting time \cite{buccinca2021trust}. Compared to Condition A, we included counterfactual explanations in Condition B and explore the effect of counterfactual explanations as a cognitive function. 

\subsubsection{Task Specifications}\label{sect:methods_cases}
\hfill\\
To investigate the effect of AI explanations on users' overreliance on `wrong' AI outputs, we utilized the trained ML models (Section \ref{sect:methods_mlmodels}) to select the cases of rehabilitation assessment for each condition. Specifically, we assigned cases with 3 `right' AI outputs and 5 `wrong' AI outputs on each condition.

\subsection{Participants \& Procedure}
\subsubsection{Participants}
\hfill\\
Seven therapists (3 male and 4 female) with an average of 12.85 years of experience in stroke rehabilitation (Table \ref{tab:demographics_summary}). In addition, we recruited ten laypersons (7 male and 3 female; 2 graduate students and 8 undergraduate students) without experience in stroke rehabilitation to compare their performance \cite{nourani2020role} and reliance on AI with expert therapists. Participants were recruited through advertisements sent to hospitals, university staff \& mailing lists, and the contacts of the research team. 

Among the seven therapists, five of them are occupational therapists whose primary roles are to help patients better engage in their daily activities. The two remaining therapists are physiotherapists who treat their patient's physical impairments from a bio-mechanical perspective. The detailed demographic information of participants is described in the Appendix (Table \ref{tab:demographics_details}).   

\begin{table}[h]
\centering
\caption{Demographics of participants (therapists and laypersons).}
\label{tab:demographics_summary}
\resizebox{\textwidth}{!}{%
\begin{tabular}{cccc|cc} \toprule
ID & Role                        & Years in the Role & Q. Tech Experience & ID  & Q. Tech Experience \\ \midrule
T1 & Occupational Therapist (OT) & 25                & 2.6 +/- 2.0        & L1  & 6.6 +/- 0.5        \\
T2 & Occupational Therapist (OT) & 5                 & 5.4 +/- 2.2        & L2  & 5.8 +/- 0.7        \\
T3 & Occupational Therapist (OT) & 10                & 4.0 +/- 2.4        & L3  & 5.0 +/- 1.7        \\
T4 & Occupational Therapist (OT) & 6                 & 3.6 +/- 2.8        & L4  & 5.6 +/- 1.5        \\
T5 & PhysioTherapist (PT)        & 17                & 3.6 +/- 2.3        & L5  & 3.0 +/- 2.1        \\
T6 & Occupational Therapist (OT) & 12                & 5.2 +/- 2.1        & L6  & 6.0 +/- 0.0        \\
T7 & PhysioTherapist (PT)        & 15                & 3.2 +/- 1.0        & L7  & 4.6 +/- 2.6        \\
   &                             &                   &                    & L8  & 5.0 +/- 1.3        \\
   &                             &                   &                    & L9  & 5.2 +/- 1.9        \\
   &                             &                   &                    & L10 & 5.2 +/- 1.8       \\ \bottomrule
\end{tabular}%
}
\end{table}

To understand the participant's background in technology, we asked them to respond to a set of technical experience questions, which were based on survey questions designed by the Center for Research and Education on Aging and Technology Enhancement (CREATE) \cite{czaja2006factors}. Each participant rated his or her experience with diverse recent technologies (i.e. computer/laptop, activity tracker, virtual voice assistant, unmanned convenient store, autonomous vehicle) on a 7-point scale (1 = strongly disagree, 2 = disagree, 3 = somewhat disagree, 4 = neutral, 5 = somewhat agree, 6 = agree, 7 = strongly agree.  A low score on technology experience (e.g. 1.0) indicates that a participant barely has experience with recent technologies. Overall, therapists have diverse levels of experience with recent technologies with an average score of 3.94 out of 7.0 and laypersons have a slightly higher average score of 5.2 out of 7.0.

\subsubsection{Procedure}
\begin{figure*}[htp]
\centering
  \includegraphics[width=\textwidth]{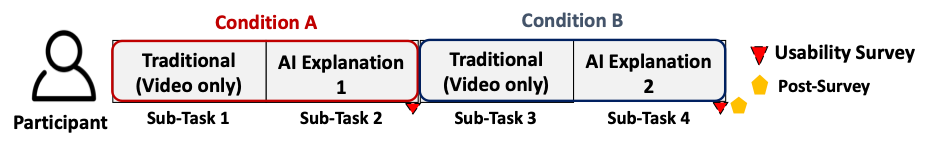}
\caption{The overall procedure of the user study: a participant completed rehabilitation assessment tasks using the AI with salient features (Condition A) and the AI with salient features and counterfactual explanations (Condition B). In each condition, the participant first completed the initial assessment after reviewing only a video and then provided the final assessment after reviewing the AI outputs and explanations. When the participant completed the tasks on each condition, the participant completed the usability questionnaires on each condition. At the end of the study, the participant completed the overall, post-survey about their preferences.}\label{fig:study-procedure}
\end{figure*}

\hfill\\
The study was conducted online. After a participant completed the informed consent form that was approved by the Institutional Review Board, the participant went through the tutorial on rehabilitation assessment and the study procedure. Each participant was randomly assigned to either first use the AI with salient feature analysis \textbf{(Condition A - Features)} and then AI with salient features and counterfactual explanations \textbf{(Condition B - Countfacts)} or vice-versa. Each condition involves two sub-tasks. Specifically, we asked the participant to (a) first provide their initial assessment (Figure \ref{fig:IntVideo}) without AI outputs and explanations and (b) then finalized the assessment after reviewing AI outputs and explanations to understand the effect of reviewing AI outputs and explanations (Figure \ref{fig:study-procedure}). In each condition, a participant was required to perform 8 decision-makings on rehabilitation assessment after reviewing post-stroke survivor's exercises. The sub-tasks of each condition were counterbalanced and the order of the two conditions and the presentations of post-stroke videos were randomized. After completing assessment tasks on each condition, the participant responded to the usability questions. After finishing all tasks on two conditions, the participant filled out the overall preference questionnaire. All participants received a fixed compensation for their participation in the study.

\subsection{Data Analysis Metrics}
We analyzed two systems (i.e. AI with salient feature analysis and AI with counterfactual explanations) using the following metrics: 1) performance and 2) participants' agreement level on rehabilitation assessment tasks, 3) counts of `right' and `wrong' decisions (including overreliance), 4) the duration of decision-making tasks, and 5) usability questionnaires \cite{lai2021towards,lee2021human,bussone2015role}. 

\subsubsection{Performance}
\hfill\\
One of the most commonly used metrics on human-AI collaborative decision-making tasks is performance, measuring the percentage of correctly making decisions on instances \cite{lai2021towards}. In this study, we utilized the annotations of a therapist from the dataset \cite{lee2019learning} as ground truths and evaluate participants' performance on decision-making tasks before/after reviewing AI outputs and explanations.

\subsubsection{Agreement Level}
\hfill\\
Most medical diagnoses rely on standardized guidelines \cite{sullivan2011fugl,early1987treatment,jankovic2008parkinson}. However, clinicians can be biased in their decision making and expert disagreement is prevalent in medical decision-making tasks
\cite{knapp1999disagreement,krause2018grader,barnett2019comparative}. Thus, we also analyzed the agreement level of participants' decisions before/after reviewing AI outputs and explanations.

\subsubsection{Counts of `right' and `wrong' decisions}
\hfill\\
In addition to the performance and agreement level, we analyzed the counts of `right' and `wrong' decisions by participants to further analyze their overreliance on `wrong' AI outputs. Also, we measured the count of (1) agreeing with `right' AI outputs, (2) rejecting `wrong' AI outputs, (3) agreeing with `wrong' AI outputs (i.e. overreliance), and (4) rejecting `right' AI outputs for further analysis. In addition, we analyzed the number of times when AI explanations assisted participants to change and make `right' or `wrong' decisions. 

\subsubsection{Duration of Decision Making}
\hfill\\
Our web interface measured the estimated duration of each decision-making by asking the participants to indicate their starting point of a decision-making task on the interface. 

\subsubsection{Usability Questionnaires}
\hfill\\
We also utilized participants' self-reported, subjective responses on usability aspects of the systems with salient feature analysis and counterfactual explanations, building upon previous research of human-AI collaborative decision-making in health \cite{lai2021towards,lee2020co,cai2019human}. Specifically, these usability aspects include (1) Useful, (2) Insight, (3) Effort, (4) Transparent, (5) Trust, (6) Frustration, (7) UsageIntent, (8) AIPotential and (9) Preference between two interfaces as follows:

\begin{itemize}
    \item Useful: \textit{``The system provided useful information to understand patient's performance for assessment''} \cite{cai2019human,lee2020co}.
    \item Insight: \textit{``The system provided new insights on patient's performance for assessment''} \cite{lee2020co}.
    \item LessEffort: \textit{``The system helped me think through and complete the assessment tasks with less effort''} based on the effort dimension of the NASA-TLX \cite{hart1988development}
    \item Reliance: \textit{``I relied on assessment scores \& analysis from the system for my final assessment''} 
    \item Transparent: \textit{``The system was transparent about why it provided a particular assessment score''} 
    \item Trust: \textit{``I can trust the provided assessment scores or/and analysis from the system''}    
    \item Frustration: \textit{``I was insecure, discouraged, and stressed while using the system''} based on the frustration dimension of the NASA-TLX \cite{hart1988development}
    \item UsageIntent: \textit{``I would use this system to understand and assess patient's exercise performance in practice''} \cite{cai2019human,lee2020co}.
    \item AIPotential: \textit{``I think AI, data-driven tool can improve rehabilitation assessment''}
    \item Preference between two interfaces: participants rated on a 7-point scale ranging from 1 (totally Condition A), 2 (much more Condition A than B), 3 (slightly more Condition A than B), 4 (neutral), ..., 7 (totally Condition B) \cite{cai2019human,lee2020co}.
\end{itemize}

All questionnaires were rated on a 7-point scale except for the trust, which was rated on a 100-point scale.

\section{Results}
Throughout this paper, we refer to the outcomes of participants, who reviewed the videos without AI outputs and explanations as \textbf{\textit{``Human''}} and those 
, who reviewed the videos with AI outputs and explanations as \textbf{\textit{``Human + AI''}}. Also, we refer the Condition A as \textbf{\textit{``Features''}}, in which participants use the AI with salient feature analysis and the Condition B as \textbf{\textit{``Counterfacts''}}, where participants use the AI with salient features and counterfactual explanations. 

For the performance and agreement level metrics, we analyzed the differences in outcomes between \textbf{\textit{``Human''}} and \textbf{\textit{``Human + AI''}} over two conditions. For the counts of \textit{`right'} and \textit{`wrong'} decisions, duration of decision makings, and usability questionnaires, we compared the outcomes of two conditions using AI with salient feature analysis and counterfactual explanations respectively. 

In the following section, we reported the descriptive statistic of each metric and conducted the Wilcox significant tests using data from therapists and laypersons respectively. If outcomes from therapists and laypersons have the same trends, we also described the overall outcomes of data analysis metrics.

\subsection{Performance}
\begin{figure*}[htp]
\centering
  \includegraphics[width=1.00\textwidth]{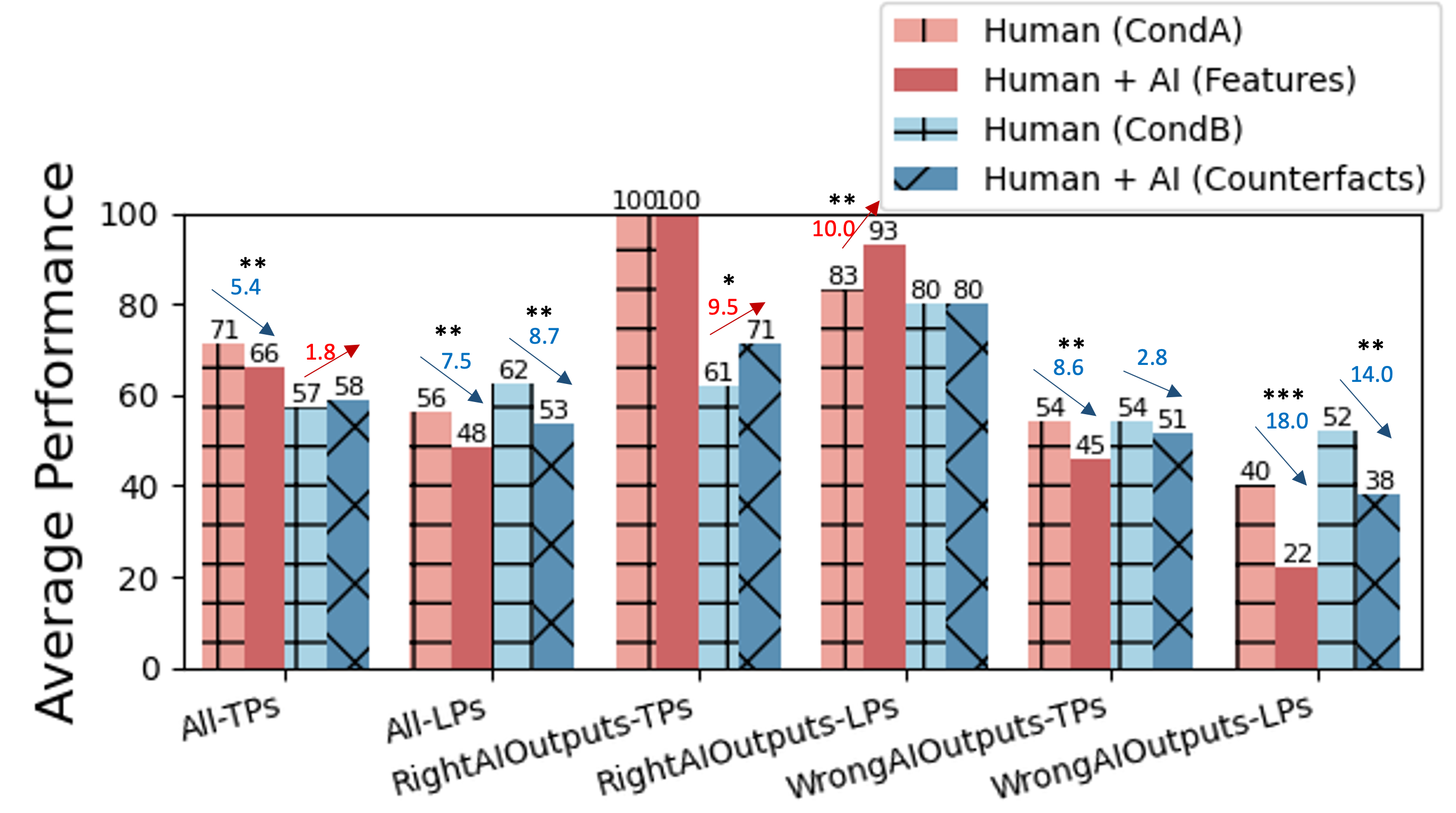}
\caption{Performance of rehabilitation assessment tasks by therapists (TPs) and laypersons (LPs) using AI with Features and Counterfactuals for (1) all cases, (2) cases with `right' AI outputs, and (3) cases with `wrong' AI outputs. Although participants' performances improved after reviewing `right' AI outputs with both salient feature analysis and counterfactual explanations, their performance reduced after reviewing `wrong' AI outputs.  Counterfactual explanations assisted participants to have lower degraded performance than salient feature analysis. *, **, and *** indicate 90\%, 95\%, and 99\% statistical significance levels.}\label{fig:results-performance-comp}
\end{figure*}

Figure \ref{fig:results-performance-comp} summarizes the average performance (i.e. F1-score) of rehabilitation assessment tasks by therapists (TPs) and laypersons (LPs) respectively. 

Overall, therapists' and laypersons' human + AI team performance with both salient feature analysis and counterfactual explanations were lower than their human alone performance ($p < 0.05$) except for a marginal improvement of therapists' human + AI team performance. For further analysis, we analyzed the performances of therapists and laypersons using the cases with `right' or `wrong' AI outputs. When `right' AI outputs were presented to therapists and laypersons, therapists' human + AI team performance with counterfactual explanations and laypersons' human + AI team performance with salient features were higher than their human alone performance ($p < 0.1$ and $p < 0.05$ respectively). However, when `wrong' AI outputs were presented to the therapists and laypersons, their human + AI team performance with salient features or counterfactual explanations was decreased. Compared to the therapists' performances, laypersons' performances were degraded significantly. Also, we found that therapists' and laypersons' human + AI team performances with salient features (i.e. 8.6 and 18.0 F1-scores respectively) led to higher performance degradation than their performance with counterfactual explanations (i.e. 2.8 and 14.0 F1-scores respectively).

\subsection{Agreement Level}
\begin{figure*}[htp]
\centering
  \includegraphics[width=0.9\textwidth]{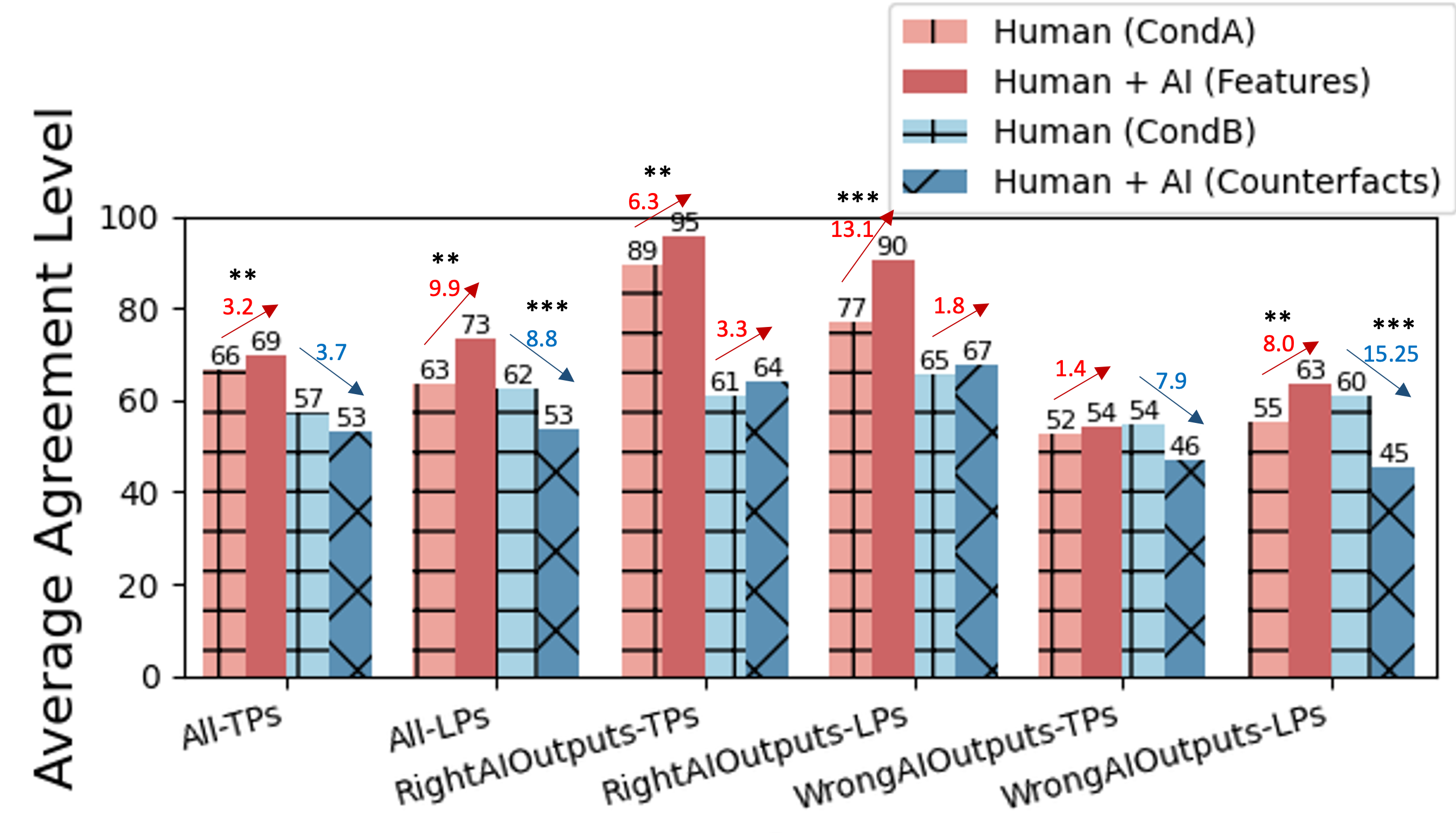}
\caption{Agreement level of rehabilitation assessment tasks by therapists (TPs) and laypersons (LPs) using AI with Features and Counterfactuals for (1) all cases, (2) cases with `right' AI outputs, and (3) cases with `wrong' AI outputs. After reviewing `right' AI outputs with salient feature analysis or counterfactual explanations, both TPs and LPs increased their agreement levels. After reviewing `wrong' AI outputs, both TPs and LPs using salient features increased their agreement levels while TPs and LPs using counterfactual explanations decreased their agreement levels. *, **, and *** indicate 90\%, 95\%, and 99\% statistical significance levels.}\label{fig:results-agreement-comp}
\end{figure*}

Figure \ref{fig:results-agreement-comp} summarizes an average agreement level (e.g. F1-score) of rehabilitation assessment tasks by therapists (TPs) and laypersons (LPs) respectively.

Overall, both TPs and LPs increased their agreement level when they reviewed AI outputs (Human + AI) of salient features ($p < 0.05$) and decreased their agreement level when they reviewed AI outputs of counterfactual explanations. 

For the cases with `right' AI outputs, both TPs and LPs achieved higher agreement levels with statistical significance ($p<0.05$ and $p < 0.01$ respectively) when they used salient features. Also, they achieved higher agreement levels without significance when they used counterfactual explanations. For the cases with `wrong' AI outputs, they increased their agreement levels (i.e. by 1.4 F1-score for TPs; by 8.0 F1-score for LPs) when they reviewed salient features. In contrast, they decreased their agreement levels when they reviewed counterfactual explanations. Similar to the performance metrics, an agreement level of TPs using counterfactual explanations (counterfacts) led to lower degradation of the agreement level (i.e. -7.9\% F1-score) than that of LPs using counterfacts (i.e. a $-15.25\%$ F1-score, $p<0.01$).

\subsection{Counts of `Right' and `Wrong' Decisions}
\begin{figure*}[htp]
\centering
  \includegraphics[width=1.0\textwidth]{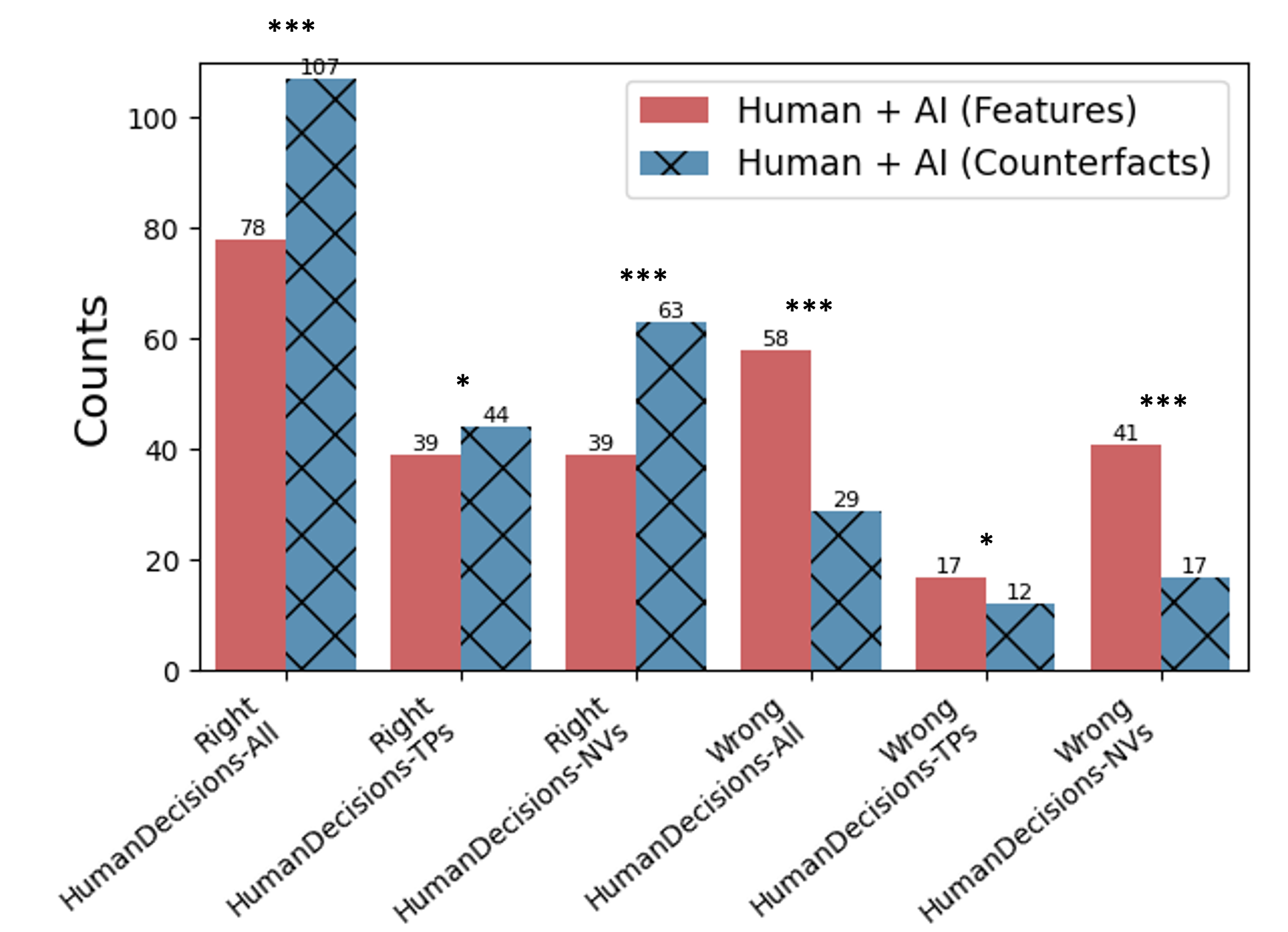}
\caption{The counts of `right' and `wrong' decisions by all participants (All), therapists (TPs), and laypersons (LPs). Counterfactual explanations assisted the participants to increase their `right' decisions and reduce their `wrong' decisions compared to the salient feature analysis. * and *** indicates 90\% and 99\% statistical significance levels.}\label{fig:results-reliance-all}
\end{figure*}

Figure \ref{fig:results-reliance-all} summarizes the counts of participants' `right' and `wrong' decisions. Overall, the human + AI team with counterfactual explanations by therapists (TPs) and laypersons (LPs) had more cases of `right' decisions than the human + AI team with salient feature analysis: 21\% (29 out 136) from all participants ($p < 0.01$), 8\% (5 out of 56) from TPs ($p < 0.1$), and 30\% (24 out of 80) from LPs ($p < 0.01$). In addition, the human + AI team with counterfactual explanations had fewer cases of `wrong' decisions than the human + AI team with salient feature analysis: 21\% (29 out 136) from all participants ($p < 0.01$), 8\% (5 out of 56) from TPs ($p < 0.1$), and 30\% (24 out of 80) from LPs ($p < 0.01$).

\begin{figure}[tp!]
\centering
\begin{subfigure}[t]{.49\columnwidth}
\centering
  \includegraphics[width=1.0\columnwidth]{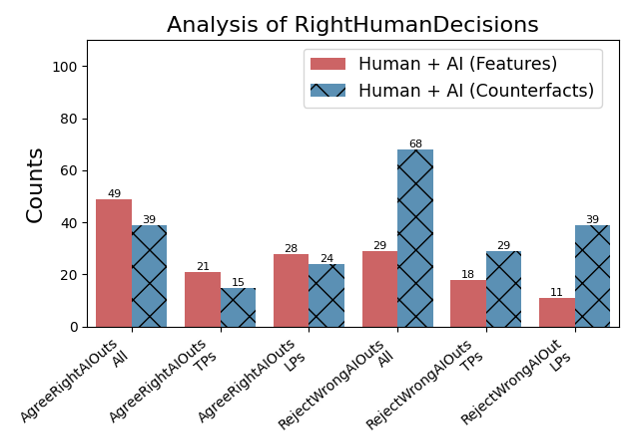}
  \caption{}
  \label{fig:results-reliance-right}
\end{subfigure}
\begin{subfigure}[t]{.49\columnwidth}
\centering
  \includegraphics[width=1.0\columnwidth]{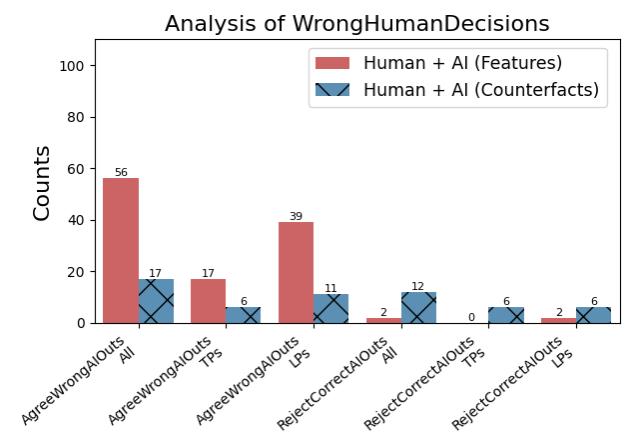}
  \caption{}
  \label{fig:results-reliance-wrong}
\end{subfigure}
\caption{Detailed analysis of (a) `right' and (b) `wrong' human decisions by all participants (All), therapists (TPs), and laypersons (LPs): Counterfactual explanations assisted the participants to increase the number of `right' decisions on rejecting `wrong' AI outputs and reduce the number of `wrong' decisions on agreeing `wrong' AI outputs.}~\label{fig:results-reliance-detailed}
\end{figure}

For the detailed analysis, we analyzed the number of `right' and `wrong' decisions of TPs and LPs by (1) agreeing with 'right' AI outputs, (2) rejecting 'wrong' AI outputs, (3) agreeing with `wrong' AI outputs, and (4) rejecting `right' AI outputs (Figure \ref{fig:results-reliance-detailed}). The human + AI team with counterfactual explanations had more cases of rejecting `wrong' AI outputs and fewer cases of agreeing with `wrong' AI outputs than the human + AI team with salient features:
by 19\% (11 out of 56) from TPs and by 35\% (28 out of 80) from LPs.
In addition, the human + AI team with Counterfacts had fewer cases of agreeing with `right' AI outputs and more cases of rejecting `right' AI outputs than the human + AI team with Features: by 10\% (6 out of 56) from TPs and by 5\% (4 out of 80) from LPs.

\subsection{Duration of Decision Making}
The participants took an average of 57 seconds (All), 49 seconds (TPs), and 63 seconds (LPs) using the system with salient feature analysis and an average of 75 seconds (All), 70 seconds (TPs), and 80 seconds (LPs) using the system with counterfactual explanations to complete a single decision-making task. Overall, the system with counterfactual explanations requires an average of 18 more seconds (All), 21 more seconds (TPs), and 17 more seconds (LPs) than the system with salient feature analysis on a decision-making task.

\subsection{Usability Questionnaires}
\begin{figure*}[htp]
\centering
  \includegraphics[width=\textwidth]{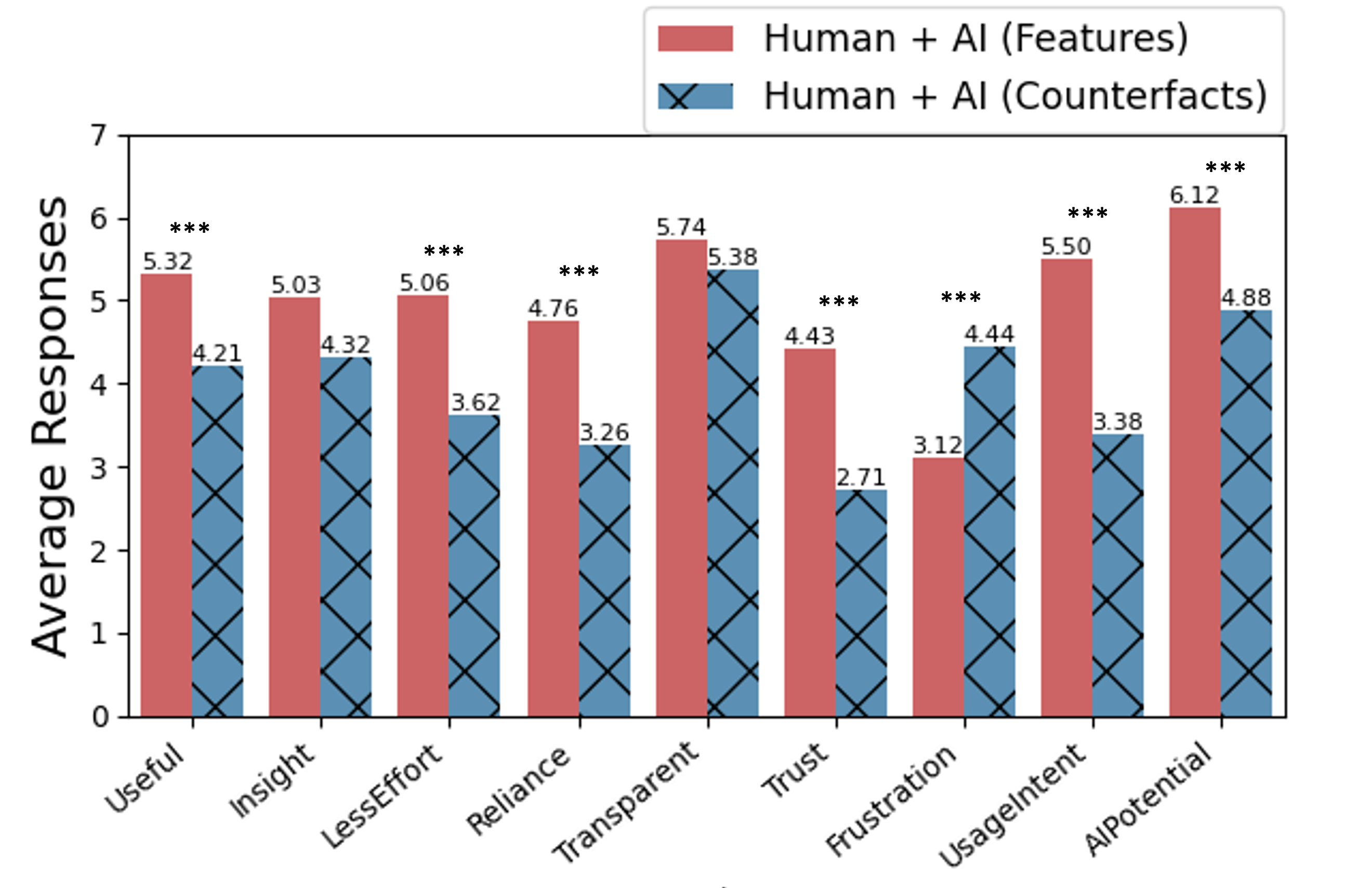}
\caption{Participants' usability responses on the system with (1) salient feature analysis (Features) and (2) counterfactual explanations (Counterfacts). Overall, participants expressed higher usage intent and potential with the system with salient feature analysis. They considered that the system with salient feature analysis is considered to be more useful, provides more insights without significance, requires less effort on assessment, is more reliable, transparent without significance, more trustful, and less frustrating. *** indicates 99\% statistical significance level.}\label{fig:results-usability-all}
\end{figure*}

Figure \ref{fig:results-usability-all} summarizes the usability responses by the participants using 
the system with (1) salient feature analysis (Features) and (2) counterfactual explanations (Counterfacts). 

The participants (both therapists and laypersons) considered that the system with Features is more useful ($p < 0.01$), provides more insights without statistical significance, requires less effort ($p < 0.01$), more reliable ($p < 0.01$), more transparent without statistical significance, more trustful ($p < 0.01$), less frustrating ($p < 0.01$). Overall, they both expressed higher usage intent ($p <0.01$) and higher potential ($p<0.01$) of the system with Features than the system with Counterfacts. 

For the post-survey on the preference question, Table \ref{tab:usability_preference} describes that there are 7 participants, who preferred the system with salient feature analysis (6 totally; 1 much more; 1 slightly more), 7 participants, who preferred the system with counterfactual explanations (2 totally; 3 much more; 3 slightly), and 1 neutral. 

Participants preferred to use the system with salient feature analysis as its visualization is \textit{``faster to read and process information''} (TP1) than textual, counterfactual explanations even if the other system. Other participants preferred the system with counterfactual explanations because they considered that these explanations assisted to \textit{``provide a second view to help assessment''} (TP 2) and \textit{``confirm any doubt during the assessment''} (TP 6).

\begin{table}[htp]
\caption{Participants' preferences on the system: overall, there are 7 participants, who preferred version A, salient feature analysis (6 totally, 1 much more, 1 slightly more), 7 participants, who preferred version B, counterfactual explanations (2 totally,  3 much more, 3 slightly more), and 1 neutral.}
\label{tab:usability_preference}
\resizebox{\textwidth}{!}{%
\begin{tabular}{cccccccc} \toprule
UserType & 
  \begin{tabular}[c]{@{}c@{}}(1) Totally\\ version A\end{tabular} &
  \begin{tabular}[c]{@{}c@{}}(2) Much more \\version A than B\end{tabular} &
  \begin{tabular}[c]{@{}c@{}}(3) Slightly more\\ version A than B\end{tabular} &
  (4) Neutral &
  \begin{tabular}[c]{@{}c@{}}(5) Slightly more\\ version B  than A\end{tabular} &
  \begin{tabular}[c]{@{}c@{}}(6) Much more\\ version B than A\end{tabular} &
  \begin{tabular}[c]{@{}c@{}}(7) Totally\\ version B \end{tabular} \\ \\ \midrule
Therapists & 2 & 1 & 1 & 0 & 1 & 2 & 1 \\ \midrule
Laypersons    & 4 & 0 & 0 & 1 & 2 & 1 & 1 \\ \midrule
Overall    & 6 & 1 & 1 & 1 & 3 & 3 & 2 \\ \bottomrule
\end{tabular}%
}
\end{table}

\section{Discussion}
In this section, we discussed the potential benefits and limitations of two types of AI explanations (i.e. salient feature and counterfactual), their effects on domain experts and laypersons' decision-making, and suggestions for more effect human-AI collaborative, clinical decision-making.

\subsection{Effects of Salient Feature \& Counterfactual Explanations for Overreliance on AI}
AI explanations have been considered as an important communication medium to realize effective human-AI collaborative decision-making tasks \cite{lai2021towards,lee2021human}. In contrast to prior research that describes the improved performance of humans with AI explanations \cite{lai2019human,lee2020co}, our results demonstrated that the human + AI team with both salient feature and counterfactual explanations performed worse than the human alone (Figure \ref{fig:results-performance-comp}). 

According to the further analysis of the cases with `right' and `wrong' AI outputs, the presentation of AI outputs and explanations has different effects on the human + AI team performance (Figure \ref{fig:results-performance-comp}). Specifically, when `right' AI outputs are presented, the human + AI team with both salient feature and counterfactual explanations performed better than the human alone. However, the human + AI team with salient feature explanations and counterfactual explanations performed worse than the human alone. 

Compared to the human + AI team with salient feature explanations, the human + AI team with counterfactual explanations supported both therapists and laypersons to have more 'right' decisions (21\%: 29 out of 136) and fewer 'wrong' decisions (21\%: 29 out of 136) (Figure \ref{fig:results-reliance-all}). Overall, our findings indicate user's overreliance on `wrong' AI outputs, which follows the previous studies that describe salient feature explanations increases user's overreliance on the AI model \cite{wang2021explanations,bansal2021does}. In addition, our results show that counterfactual explanations performed better than salient feature explanations to assist reduce therapists' and laypersons' overreliance on AI.

When it comes to the agreement level, our results showed that the human + AI team with salient feature analysis led to an increase in the agreement level on the cases with `right' and `wrong' AI outputs (Figure \ref{fig:results-agreement-comp}). However, we found that the increase in agreement level does not necessarily indicate a positive performance improvement. Specifically, the counts of participants' `right' and `wrong' decisions (Figures \ref{fig:results-reliance-all} and \ref{fig:results-reliance-detailed}) indicated that the human + AI team with salient feature analysis had more `wrong' decisions while having a lower number of rejecting `wrong' AI outputs and a higher number of agreeing `wrong' AI outputs than the human AI team with counterfactual explanations. Taken together, our findings show that counterfactual explanations can serve as a cognitive forcing function \cite{buccinca2021trust} that assists the users in analytically reviewing AI explanations and reducing their overreliance on `wrong' AI outputs.  

\subsection{Domain Experts vs Laypersons}
Among various data analysis metrics, we found that both therapists and laypersons had mostly the same outcome patterns in performance, agreement level, counts of `right' and 'wrong' decisions, duration of decision-making, and usability responses. However, our results suggest that laypersons had a higher over-reliance on AI outputs than therapists. 

Specifically, when `wrong' AI outputs were presented, laypersons had much higher performance degradation by 18.0 f1-score with salient feature explanations and 14.0 f1-score with counterfactual explanations than therapists who had a performance degradation of 8.6 f1-score with salient feature explanations and 2.8 f1-score with counterfactual explanations. In addition, this over-reliance on `wrong' AI outputs has shown more significance with laypersons than domain experts, therapists. 

\subsection{Towards Better Calibrated Trust and Evaluation Metrics on AI}
Similar to the previous research \cite{bussone2015role}, our study also shows a positive correlation between user's trust and reliance: the more a user trusts the system, the more the user is likely to over-rely on outputs of the system even when `wrong' AI outputs are presented. Although we provided the same number of `right' and `wrong' AI outputs on two systems with salient feature analysis and counterfactual explanations, participants considered that the system with salient feature analysis \textit{``is more accurate''} (TP 6) than the system with counterfactual explanations. 

Overall, participants expressed that they had a higher, self-reported trust score and a higher reliance score on the system with salient feature analysis than the system with counterfactual explanations. In particular, the trust score of the system with salient feature analysis is 73.76 out of 100 and that of the system with counterfactual explanation is 45.20 out of 100. 

As our task specification (Section \ref{sect:methods_cases}) includes 3 `right' and 5 `wrong' AI outputs, the ideal estimation of an ML performance is 0.375 (3 out of 8). The participants using the system with neither explanation exactly estimated the performance of an ML model. However, the trust score of the system with counterfactual explanation is much closer than that of the system with salient feature analysis. Thus, this finding suggests that counterfactual explanations also have the potential to assist a user in better evaluating and estimating the accuracy of an ML model. 

In addition, our findings suggest that possible gaps between users' perceived benefits and actual trustworthiness of an AI system. Relying only on subjective usability responses \cite{lai2021towards} might be limited and does not provide an appropriate understanding and evaluation on the trustworthiness of an AI system. In other words, an AI system with a higher self-reported trust score by participants does not necessarily mean that the system would achieve human + AI complementary team performance. It is important to explore a way or metrics to more accurately evaluate the trustworthiness and effectiveness of AI systems in the future.  

\subsection{Limitations}
Our results demonstrated the potential of the human + AI team with counterfactual explanations to reduce the overreliance on AI and make the users better estimate the accuracy of an AI model during human-AI collaborative decision-making using uncontrolled AI outputs and explanations that are more stochastic. However, participants had lower usage intent and expected lower potential of the AI system with counterfactual explanations as reviewing a counterfactual explanation presented in texts \textit{``could be more confusing''} (T1) and \textit{``take more effort to complete the assessment''} (T4). 

As previous research shows the higher understandability of counterfactual explanations by clinicians by including visual graphics than textual descriptions \cite{naiseh2023different}, we believe our limited scores of usability aspects on counterfactual explanations might be overcome by exploring new visualizations and human-centered design of AI explanations \cite{gomez2020vice,kaul2021improving}. 

As this work primarily focuses on exploring the effect of counterfactual explanations on user trust, the experimental designs of this work do not consider the possible effect of explanation fidelity. It is important to further investigate the effect of explanation fidelity on user trust \cite{papenmeier2019model} and overreliance. 
In addition, an additional study is required to investigate how people can effectively evaluate the performance and trustworthiness of an ML model and calibrate their trust and reliance to improve human-AI/algorithm interaction \cite{green2019principles}.

Our work also has a limitation in its generalizability as our work does not involve a large number of participants. However, such a small sample size is not unusual in similar previous works \cite{lee2020co,bussone2015role}. In addition, this study mainly explores our research question in the context of a single clinical decision-making task (i.e. rehabilitation assessment) and is limited by particular types and visualization formats of AI explanations and an ML model (i.e. a feed-forward neural network). It is required to conduct additional studies to explore other decision-making tasks and types of ML models, explanations, and visualizations \cite{gomez2020vice,kaul2021improving} for further generalization of our findings.

\section{Conclusion}
In this work, we contributed to an empirical study of analyzing the effect of the salient feature and counterfactual explanations on users' trust and reliance on AI during a human-AI collaborative clinical decision-making task (i.e. assessing post-stroke survivor's quality of motion). Our results showed that the humans + AI team with both salient feature and counterfactual explanations increased its performance on decision-making tasks only when `right' AI outputs are presented and decreased its performance when `wrong' AI outputs are presented. Our results demonstrated that counterfactual explanations assisted the participants to reduce their overreliance on `wrong' AI outputs (21 \%) compared to salient feature explanations. Also, we found that laypersons had higher performance degradation and overreliance than domain experts, therapists. Taken together, our work brings to light that providing AI explanations does not necessarily indicate improved human-AI collaborative decision-making. This work discusses the potential of counterfactual explanations to improve analytical reviews on AI outputs to better estimate AI performance and reduce overreliance on AI with the cost of cognitive burdens and other implications for improving human-AI collaborative decision-making.

\begin{acks}
The authors thank all the participants in this work for their time and valuable inputs. We also thank the anonymous reviewers for their constructive feedback. This work is supported by the Singapore Ministry of Education (MOE) Academic Research Fund (AcRF) Tier 1 grant.
\end{acks}

\bibliographystyle{ACM-Reference-Format}
\bibliography{main}

\newpage
\appendix

\begin{table}[htp!]
\centering
\caption{Guidelines to Assess Stroke Rehabilitation Exercises}
\label{tab:score-guidelines}
\resizebox{0.7\columnwidth}{!}{%
\begin{tabular}{ccl} \toprule
\textbf{\begin{tabular}[c]{@{}c@{}}Performance \\ Components\end{tabular}} & \textbf{Score} & \multicolumn{1}{c}{\textbf{Guidelines}} \\  \midrule
\multirow{3}{*}{\begin{tabular}[c]{@{}c@{}}Range of Movement \\ (ROM)
\end{tabular}
} & 0 & Does not or barely involve any movement \\
 & 1 & Less than half way aligned with an \textit{`Target'} position
 \\
 & 2 & Movement achieves an \textit{`Target'} position \\ \midrule
\multirow{3}{*}{Compensation} & 0 & Noticeable compensation in more than two joints \\
 & 1 & Noticeable compensation in a joint \\
 & 2 & Does not involve any compensations \\ \bottomrule
\end{tabular}%
}
\end{table}

\begin{table}[!htb]
    \caption{Parameters of Machine Learning Models (i.e. Feed-Forward Neural Network Models)}\label{tab:params-ml}
      \centering
      \resizebox{0.45\columnwidth}{!}{%
       \begin{tabular}{ccc} \toprule
\multicolumn{1}{l}{} & \multicolumn{2}{c}{\begin{tabular}[c]{@{}c@{}}Hidden Layers and Units / Learning Rate\end{tabular}} \\ \toprule
 & ROM & Comp \\ \midrule
E1 & (256) / 0.005 & (16, 16) / 0.01 \\  \bottomrule
\end{tabular}%
}
\end{table}

\begin{table}[htp]
\centering
\caption{Detailed Demographics of Therapists (T1 - T7) and Laypersons (L1 - L10)}
\label{tab:demographics_details}
\resizebox{\textwidth}{!}{%
\begin{tabular}{cccclc} \toprule
ID  & Gender & Age             & Q. Tech Experience & \multicolumn{1}{c}{Occupation}    & Years in the Role \\ \midrule
T1  & Female & 45 - 54   years & 2.6 +/- 2.0        & Occupational Therapist (OT) & 25                \\
T2  & Female & 25 - 34 years   & 5.4 +/- 2.2        & Occupational Therapist (OT) & 5                 \\
T3  & Female & 25 - 34 years   & 4.0 +/- 2.4        & Occupational Therapist (OT) & 10                \\
T4  & Male   & 25 - 34 years   & 3.6 +/- 2.8        & Occupational Therapist (OT) & 6                 \\
T5  & Male   & 35 - 44 years   & 3.6 +/- 2.3        & PhysioTherapist (PT)        & 17                \\
T6  & Female & 25 - 34 years   & 5.2 +/- 2.1        & Occupational Therapist (OT) & 12                \\
T7  & Male   & 35 - 44 years   & 3.2 +/- 1.0        & PhysioTherapist (PT)        & 15   \\ \midrule        
L1  & Female & 25 - 34  years  & 6.6 +/- 0.5        &            Graduate Student                &                   \\
L2  & Female & 25 - 34 years   & 5.8 +/- 0.7        &            Graduate Student                  &                   \\
L3  & Male   & 18 - 24 years   & 5.0 +/- 1.7        &            Undergraduate Student          &                   \\
L4  & Male   & 18 - 24 years   & 5.6 +/- 1.5        &            Undergraduate Student                 &                   \\
L5  & Male   & 18 - 24 years   & 3.0 +/- 2.1        &            Undergraduate Student                 &                   \\
L6  & Male   & 18 - 24 years   & 6.0 +/- 0.0        &            Undergraduate Student                 &                   \\
L7  & Male   & 18 - 24 years   & 4.6 +/- 2.6        &            Undergraduate Student                 &                   \\
L8  & Male   & 18 - 24 years   & 5.0 +/- 1.3        &             Undergraduate Student                &                   \\
L9  & Male   & 18 - 24 years   & 5.2 +/- 1.9        &            Undergraduate Student                 &                   \\
L10 & Female & 18 - 24 years   & 5.2 +/- 1.8        &            Undergraduate Student                 &                   \\ \bottomrule
\end{tabular}%
}
\end{table}

\received{January 2023}
\received[revised]{April 2023}
\received[accepted]{May 2023}

\end{document}